\documentclass[aip,jap,unsortedaddress,superscriptaddress,reprint]{revtex4-1}
\usepackage{hyperref}
\usepackage{graphicx}
\usepackage{amsmath}
\usepackage{amssymb}
\usepackage{times,mathptmx}
\usepackage{dcolumn}
\usepackage{ifthen}
\usepackage{color}
\usepackage{proof}

\newif\ifcom
\newif\ifdel
\comtrue               %
\delfalse

\begin{document}

\title{Epitaxial Growth of Spinel Cobalt Ferrite Films on MgAl$_2$O$_4$ Substrates by Direct Liquid Injection Chemical Vapor Deposition}
\author{Liming Shen}
\affiliation{University of Alabama, Center for Materials for Information Technology MINT and Dept.~of Chem, Tuscaloosa, AL 35487 USA}
\author{Matthias Althammer}
\affiliation{University of Alabama, Center for Materials for Information Technology MINT and Dept.~of Chem, Tuscaloosa, AL 35487 USA}
\author{Neha Pachauri}
\affiliation{University of Alabama, Center for Materials for Information Technology MINT and Dept.~of Chem, Tuscaloosa, AL 35487 USA}
\author{B. Loukya}
\affiliation{International Centre for Materials Science, Jawaharlal Nehru Centre for Advanced Scientific Research, Jakkur P.O., Bangalore 560064, India}
\author{Ranjan Datta}
\affiliation{International Centre for Materials Science, Jawaharlal Nehru Centre for Advanced Scientific Research, Jakkur P.O., Bangalore 560064, India}
\author{Milko Iliev}
\affiliation{Texas Center for Superconductivity and Department of Physics, University of Houston, Houston, Texas 77204-5002, USA}
\author{Ningzhong Bao}
\affiliation{State Key Laboratory of Material-Oriented Chemical Engineering, College of Chemistry and Chemical Engineering, Nanjing University of Technology, Nanjing, Jiangsu 210009, P. R. China}
\author{Arunava Gupta}
\email{agupta@mint.ua.edu}
\affiliation{University of Alabama, Center for Materials for Information Technology MINT and Dept.~of Chem, Tuscaloosa, AL 35487 USA}
\date{\today}
\begin{abstract}
The direct liquid injection chemical vapor deposition (DLI-CVD) technique has been used for the growth of cobalt ferrite (CFO) films on (100)-oriented MgAl$_2$O$_4$ (MAO) substrates. Smooth and highly epitaxial cobalt ferrite thin films, with the epitaxial relationship $\mathrm{MAO} (100)\:[001] \parallel \mathrm{CFO} (100)\:[001]$,  are obtained under optimized deposition conditions. The films exhibit bulk-like structural and magnetic properties with an out-of-plane lattice constant of $8.370\;\mathrm{\AA}$ and a saturation magnetization of $420\;\mathrm{kA/m}$ at room temperature. The Raman spectra of films on MgAl$_2$O$_4$ support the fact that the Fe$^{3+}$- and the Co$^{2+}$-ions are distributed in an ordered fashion on the B-site of the inverse spinel structure. The DLI-CVD technique has been extended for the growth of smooth and highly oriented cobalt ferrite thin films on a variety of other substrates, including MgO, and piezoelectric lead magnesium niobate-lead titanate and lead zinc niobate-lead titanate substrates.
\end{abstract}
\maketitle
\section{Introduction}
\label{Introduction}
Cobalt ferrite, CoFe$_2$O$_4$ (CFO), with the inverse spinel structure is well known to have a relatively large magnetic anisotropy, high Curie temperature of around $800\;\mathrm{K}$, moderate saturation magnetization, remarkable chemical stability, and good mechanical hardness. These excellent properties offer CFO numerous technological applications in areas such as hard pinning layers, pressure sensors, actuators, and transducers, spin filters, and drug delivery.\cite{chen_metal-bonded_1999,carey_spin_2002,hassan_structural_2007} Recently, much attention has been focused on the growth of epitaxial spinel ferrite thin films on piezoelectric substrates - a type of magnetoelectric multiferroic composite structure proposed to overcome the limited choice of single-phase multiferroic materials exhibiting coexistence of strongly coupled ferro/ferrimagnetism and ferroelectricity.~\cite{spaldin_renaissance_2005,ramesh_multiferroics:_2007} The coexistence of ferroelectric and magnetic phases in such multiferroic heterostructures produces novel properties resulting from the interfacial coupling of structural, electrical, and magnetic order parameters. Because of its large magnetostrictive coefficient, CFO has been considered as an important component for multiferroic composites that exhibit a strong coupling of the order parameters through the heteroepitaxy of the two lattices.~\cite{wang_giant_2011,zheng_three-dimensional_2004,chen_enhanced_2010} In order to couple via the interface, epitaxial growth of CFO with chemically sharp, smooth interfaces, as well as good magnetic properties, is critical.
Up to date, epitaxial CFO films have been successfully fabricated by pulsed laser deposition (PLD),~\cite{ma_robust_2010} molecular beam epitaxy (MBE),~\cite{chambers_molecular_2002} and metal organic chemical vapor deposition (MOCVD).~\cite{fujii_preparation_1995} These CFO films have been grown primarily on single crystal substrates such as perovskite SrTiO$_3$ (STO), isostructural spinel MgAl$_2$O$_4$ (MAO), rocksalt MgO, and various buffered substrates. The thickness of these films is usually in the range of several nanometers up to $400\;\mathrm{nm}$, but their magnetic properties are usually far from bulk properties and also the quality of film microstructure decreases with increasing thickness. 3D island formation at high growth temperature, large lattice mismatch and the mismatch of thermal expansion coefficients between the film and substrates have been considered as the major obstacles to achieve high quality CFO thin films.~\cite{ma_robust_2010} Thus, it remains a challenge for developing efficient synthetic techniques to fabricate epitaxial thick CFO films with near-perfect microstructure and excellent magnetic properties close to bulk CFO.
Direct liquid injection chemical vapor deposition (DLI-CVD) is an attractive method for the fabrication of multi-component stoichiometric films because of its accurate delivery of several precursors dissolved in a common solvent.~\cite{senzaki_mocvd_2000} This deposition method also has the advantages of relatively low cost, high deposition rates, industrial compatibility, and the flexibility to fabricate multilayer structures over conventional thin film processes. DLI-CVD has already been utilized to prepare atomically smooth NiFe$_2$O$_4$ (NFO) thin films at temperatures of $600-700^\circ\mathrm{C}$ with saturation magnetization values only slightly lower than bulk NiFe$_2$O$_4$.~\cite{li_growth_2011} As CFO possesses a similar spinel ferrite structure like NiFe$_2$O$_4$, DLI-CVD should also potentially enable the growth of high quality CFO thin films with bulk-like properties. In this work, we present the results of our growth studies and the structural and magnetic properties of high-quality thick CFO films ($\approx 1\;\mathrm{\mu m}$) epitaxially grown on (100)-oriented MAO substrates by DLI-CVD method. A narrow growth temperature window is found for the growth of high quality CFO thin film. Last but not least, we show that the optimized deposition parameters are suitable for the growth of thick CFO films on a variety of other substrates [(MgO, lead magnesium niobate-lead titanate (PMN-PT), and lead zinc niobate-lead titanate (PZN-PT)] with excellent structural quality.
\section{Experimental}
\label{Experimental}
The DLI-CVD setup used for the experiments has already been described in a previous report.~\cite{li_growth_2011} In our growth process, anhydrous Co(acac)$_3$ and Fe(acac)$_3$ (acac is an abbreviation for acetylacetonate) in the molar ratio of 1:2 were dissolved in N,N-dimethyl formamide (DMF) to form a clear homogeneous precursor solution. The precursor solution was then introduced into a Brooks Instrument DLI 200 vaporizer system, vaporized by heated Ar gas, followed by reaction with oxygen (O$_2$) gas on heated substrates placed in a CVD reactor, resulting in the formation of a stoichiometric CFO film on the substrate. The flow rates of Ar and O$_2$ gas were optimized to be 400 and 150 sccm, respectively, with a precursor injection rate of $9.5\;\mathrm{g/h}$. The film deposition was carried out at temperatures ranging from $550\;^\circ\mathrm{C}$ to $700\;^\circ\mathrm{C}$ for 1 h. After deposition, the films were cooled to room temperature in the reactor under flowing O$_2$. Film morphology was characterized by field emission scanning electron microscopy (FE-SEM, JEOL-7000) as well as atomic force microscopy (AFM, Veeco nanoscope-IV). To determine the crystal phase and epitaxy of the resulting films, a standard 4 circle x-ray diffraction (XRD) setup (Phillips X’ pert Pro) was used with a Cu $K\alpha$ source. Magnetic properties of the CFO films were determined via superconducting quantum interference device (SQUID) magnetometry (MPMS by Quantum Design) at temperatures $5\;\mathrm{K} \leq T \leq 350\;\mathrm{K}$ and in magnetic fields of up to $5\;\mathrm{T}$. Polarized Raman spectra were measured with $XX$, $XY$, $X'X'$, and $X'Y'$ scattering configurations using a $488\;\mathrm{nm}$ excitation source at room temperature. In these notations the first and second letters indicate the polarization of the incident and scattered light, respectively, along the cubic $X\parallel[100]$, $Y\parallel[010]$, $X'\parallel[110]$, or $Y'\parallel[-110]$ directions of the MAO substrate. Transmission electron microscope (TEM) cross sectional samples were prepared by conventional mechanical polishing and Ar ion milling. Electron diffraction and imaging were performed in a FEI TITAN$^3$ \texttrademark $\;80-300\;\mathrm{kV}$ aberration corrected transmission electron microscope.

\section{Results and Discussion}
\label{Results}
We first look into the structural quality of CFO films grown on MAO (100) substrates at temperatures ranging from 550 to 700 $^\circ\mathrm{C}$. All the other optimized growth parameters, such as flow rate and vaporization temperature of precursor solutions, flow rate of O$_2$, and the total pressure in the reactor, were kept constant for all films.
\begin{figure}[h,b,t]
  \includegraphics[width=85mm]{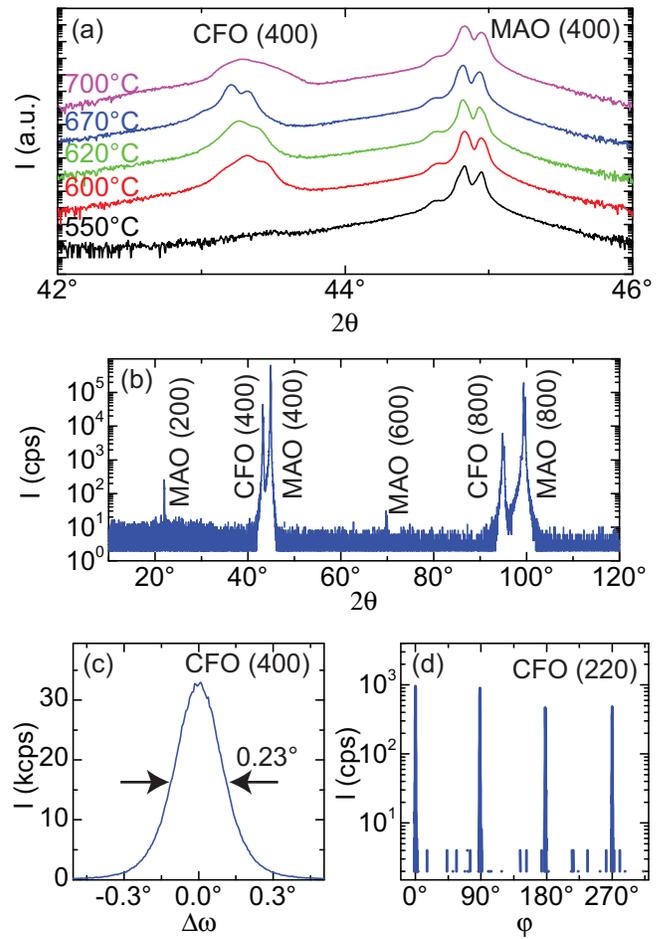}\\
  \caption[X-ray scans on MAO]{(Color online) (a) X-ray diffraction $2\theta-\omega$ scans of CFO film grown on (100) oriented MAO at $550\;^\circ\mathrm{C}$ (black), $600\;^\circ\mathrm{C}$ (red), $620\;^\circ\mathrm{C}$ (green), $670\;^\circ\mathrm{C}$ (blue), and $700\;^\circ\mathrm{C}$ (magenta) . (b) wide-range $2\theta-\omega$ scan, (c) rocking curve of the CFO (400) reflection, and (d) phi ($\varphi$)-scan of the CFO (220) reflection for a CFO film grown at $670^\circ\;\mathrm{C}$. The narrow rocking curve with a FWHM of $0.23^\circ$ is indicative of the high structural quality. The phi-scan exhibits a 4-fold symmetry indicating a cube-on-cube growth of CFO on MAO.}
  \label{figure:XRD_MAO}
\end{figure}
Figure~\ref{figure:XRD_MAO}(a) shows the normal Bragg XRD spectra near the (400) reflection of the MAO substrate of CFO films grown at different deposition temperatures. The CFO grown at $550\;^\circ\mathrm{C}$ exhibits no film reflections and thus is amorphous, which is probably due to the low deposition temperature, which leads to a low crystallization rate of the film on the substrate. CFO films grown between 600 and $670\;^\circ\mathrm{C}$ exhibit a single-phase spinel structure, without any additional peaks from impurity phases. The characteristic (400) reflection of the CFO films become narrower and sharper as the deposition temperature increases, indicating that the epitaxy and microstructure improves steadily. The appearance of a sharp $K\alpha$ peak splitting and high intensity of the (400) film reflection at $2\theta\approx 43.25^\circ$ for the CFO film grown at $670\;^\circ\mathrm{C}$ [Fig.~\ref{figure:XRD_MAO}(a)] indicates a highly epitaxial growth for this deposition temperature. For higher growth temperatures the quality of the films starts to degrade, as evident from the disappearance of the sharp $K\alpha$ peak splitting for a film grown at $700\;^\circ\mathrm{C}$.

The film grown at $700\;^\circ\mathrm{C}$ shows a very rough cracked surface with some large area film peeling. The rough surface can be caused by a faster deposition rate at higher temperature causing a higher density of defects. 
These films can easily peel off from the substrate surface during the cooling process due to stress caused by the difference in thermal expansion coefficients of film and substrate. Therefore, spinel CFO films with good epitaxy and texture can only be prepared within a narrow growth temperature window around $670\;^\circ\mathrm{C}$ in our DLI-CVD setup. The thickness of the films can be directly controlled by the deposition time.

From the XRD results in Fig.~\ref{figure:XRD_MAO}(a) we extracted the following out-of-plane lattice constants for CFO thin films grown at 600, 620, and $670\;^\circ\mathrm{C}$: $8.352\;\mathrm{\AA}$, $8.360\;\mathrm{\AA}$, and $8.370\;\mathrm{\AA}$, respectively. Although all the CFO films are rather thick ($\approx1\;\mathrm{\mu m}$), the lattice constants of these samples are smaller than the bulk value ($8.377\;\mathrm{\AA}$), suggesting an in-plane tensile stress in the films. As the lattice constant of MAO ($8.09\;\mathrm{\AA}$) is smaller, the occurrence of an in-plane tensile stress is unexpected. Gao \textit{et al.}~\cite{gao_switching_2009} found that the unexpected in-plane tensile stress in CFO/SrRuO$_3$ heterostructure is strongly correlated with the density of the interface misfit dislocations, which are formed at the deposition temperature to relieve the stress caused by the lattice mismatch. Besser \textit{et al.}~\cite{besser_stress_1971} proposed a stress model for heteroepitaxial magnetic oxide films based on the different thermal expansion rates between the substrate and the film during the cooling process. The residual stress after cooling the sample to room temperature is proportional to the difference in the thermal expansion coefficient between the film and substrate and is not related to the bulk lattice parameters. A tensile stress on CFO can be created because CFO ($14.9 \times 10^{-6}\;\mathrm{K^{-1}}$) ~\cite{li_growth_2011} has a larger thermal expansion coefficient than that of MAO ($9.5 \times 10^{-6}\;\mathrm{K^{-1}}$),~\cite{gao_switching_2009} resulting in a faster contraction rate in CFO than in MAO when cooling the samples from the deposition temperature to room temperature. From the reduction of the tensile stress in our samples for increasing temperature, we conclude that the difference in thermal expansion is not the cause for stress, but more likely the formation of misfit dislocations is dominant. 

Figure~\ref{figure:XRD_MAO}(b) shows a $2\theta-\omega$ XRD scan over a larger $2\theta$ range for the CFO sample deposited at $670\;^\circ\mathrm{C}$, exhibiting only (h00) reflections of substrate and film. No secondary phases are visible in this full range XRD scan. These finding support that we are able to grow single phase CFO films by DLI-CVD.

To analyze the film mosaic spread we investigated the width of the rocking curve ($\omega$-scan) [Fig.~\ref{figure:XRD_MAO}(c)] of the CFO (400) reflection for the sample grown at $670\;^\circ\mathrm{C}$. The measured full width at half maximum (FWHM) is $0.22^\circ$, which is comparable to our results on NFO films.~\cite{li_growth_2011}

To determine the epitaxial relationship between film and substrate, we carried out XRD phi($\varphi$)-scans at the (220) reflection of the CFO film and MAO substrate. The results for the CFO(220) reflection are shown in Fig.~\ref{figure:XRD_MAO}(d). The $\varphi$)-scan exhibits four sharp reflections, $90^\circ$ apart, confirming the cubic symmetry of the CFO films grown at $670\;^\circ\mathrm{C}$. Furthermore, we conclude that the epitaxial relationship is $\mathrm{MAO} (100)\:[001] \parallel \mathrm{CFO} (100)\:[001]$.

\begin{figure}[h,b,t]
  \includegraphics[width=85mm]{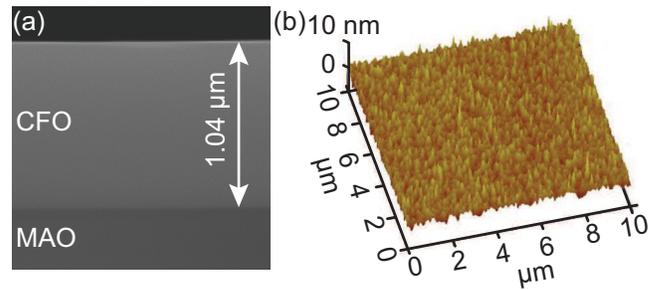}\\
  \caption[SEM and AFM for CFO on MAO]{(Color online) (a) Cross-sectional SEM image and (b) AFM image of a CFO film grown on (100)-oriented MAO at $670\;^\circ\mathrm{C}$. Both images show that the film has a very low surface roughness. From the AFM measurement we obtain a RMS roughness of $1.1\;\mathrm{nm}$.}
  \label{figure:SEM_AFM}
\end{figure}

The morphology and surface roughness of the synthesized films has been characterized by SEM and AFM. Figure \ref{figure:SEM_AFM}(a) shows the cross-section SEM image of a CFO film deposited at $670\;^\circ\mathrm{C}$. The sample exhibits a clear and sharp interface between the CFO film and MAO substrate. The thickness of the film is measured to be $1.04\;\mathrm{\mu m}$. The film appears very smooth and dense, without any bulk cracks observed through the entire cross section. Figure \ref{figure:SEM_AFM}(b) is the 3D surface plot of a $10\times 10\;\mathrm{\mu m^2}$ AFM image of the sample surface, exhibiting a root mean square (RMS) roughness of $1.1\;\mathrm{nm}$. No cracks are observed over the entire film surface.

The film cation stoichiometry was determined by energy-dispersive x-ray Spectroscopy (EDS). The EDS analysis results yields a Fe/Co ratio close to the expected value of 2.0, indicating a stoichiometric CFO film without contaminations.

\begin{figure}[h,b,t]
  \includegraphics[width=85mm]{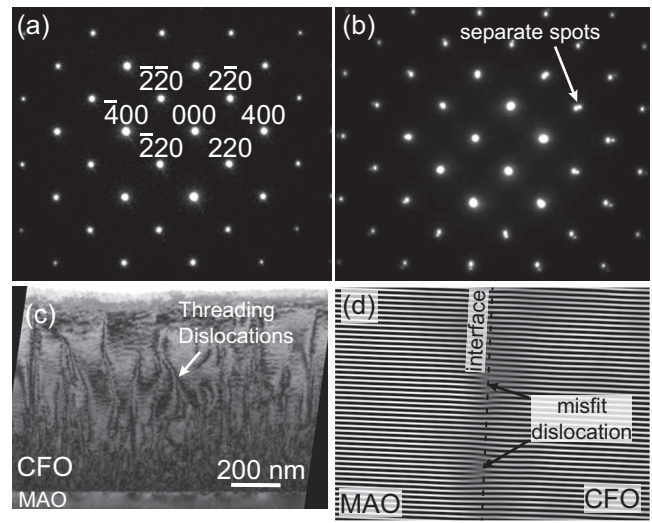}\\
  \caption[TEM CFO on MAO]{(Color online) Selected area electron diffraction pattern (a) from cobalt ferrite film only; (b) from both the film and the MAO substrate for a CFO film grown on MAO(100) at $670\;^\circ\mathrm{C}$. These images point to a relaxed growth of CFO on MAO consistent with the epitaxial relationship determined from XRD measurements: $\mathrm{MAO} (100)\:[001] \parallel \mathrm{CFO} (100)\:[001]$. (c) The bright field TEM image with $g=\langle 400 \rangle$ reveals several misfit dislocations within the CFO thin film. (d) Fourier filtered image showing formation of misfit dislocations at film-substrate interface.}
  \label{figure:TEM}
\end{figure}

Transmission electron microscopy (TEM) images and selected area electron diffraction (SAED) patterns have been recorded to further examine the epitaxial growth and microstructure of the CFO film grown at $670\;^\circ\mathrm{C}$. SAED patterns taken along the $[001]$ direction and diffraction spots from different lattice planes of CFO film are indexed in Fig.~\ref{figure:TEM}(a). In Figure~\ref{figure:TEM}(b), two sets of diffraction spots, originating from the CFO film and MAO (100) substrate, overlap with each other at lower order diffractions and slightly separate at higher order diffractions. This confirms the relaxed growth of CFO on MAO under these deposition conditions. Moreover, these results confirm the epitaxial relationship found in XRD experiments [c.f. Fig.~\ref{figure:XRD_MAO}(d)]: $\mathrm{MAO} (100)\:[001] \parallel \mathrm{CFO} (100)\:[001]$.

The bright field TEM ($g=\langle 400 \rangle$) image [Fig.~\ref{figure:TEM}(c),(d)] of the sample reveals the microstructure of the film in detail. Threading dislocations and dark diffused contrast areas are observed in these images. These dislocations are likely generated during the thin film strain relaxation process caused by lattice mismatch and/or thermal expansion difference between the film and substrate. Some dark diffused contrast may result from anti-phase domains or cation ordering, as suggested by Datta \textit{et al.}.~\cite{datta_formation_2010,datta_structural_2012} Figure~\ref{figure:TEM}(d) shows the Fourier filtered image of a interface area. At the interface of CFO and MAO we observe some misfit dislocations. Nevertheless, these results demonstrate that using the optimized set of deposition parameters we obtain CFO films with a low density of dislocations.

\begin{figure}[h,b,t]
  \includegraphics[width=85mm]{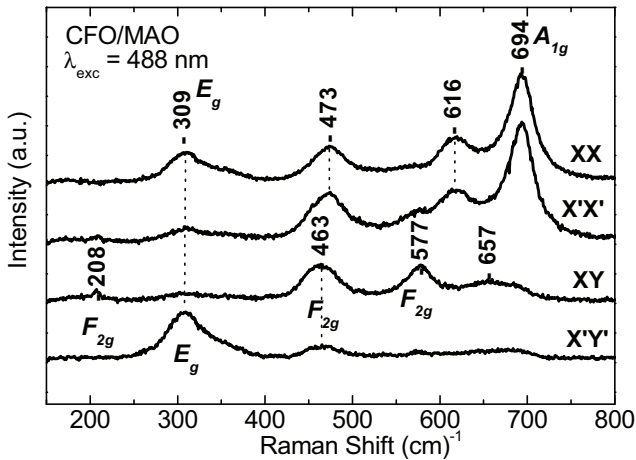}\\
  \caption[Raman for CFO on MAO]{Polarized Raman spectra obtained for a CFO film on MAO substrate grown at $670\;^\circ\mathrm{C}$ using a 488 nm excitation at room temperature. Additional lines of $A_{1g}$ symmetry are observed at $473\;\mathrm{cm^{-1}}$ and $616\;\mathrm{cm^{-1}}$. This suggests an ordered distribution of Co$^{2+}$ and Fe$^{3+}$ on the octahedral sites.}
  \label{figure:RAMAN}
\end{figure}

Figure \ref{figure:RAMAN} shows the polarized Raman spectra of a CFO film grown on MAO(100) substrate at $670\;^\circ\mathrm{C}$ with a $488\;\mathrm{nm}$ excitation at room temperature. The numbers of Raman lines and their positions are consistent with those of previous studies of CFO samples in the forms of crystals, powders, and thin films.~\cite{foerster_poisson_2012,liao_temperature_2012,wang_flux_2006,yu_cation_2002} The apparent differences between $XY$ and $X'Y'$ spectra and between $XX$ and $X'X'$ spectra provide unambiguous evidence that the film is of excellent epitaxial quality. The number of the Raman lines exceeds that expected for a normal spinel or an inverse spinel with Co/Fe disorder at the B-sites, where only five Raman modes ($A_{1g} + E_{g} + 3 F_{2g}$) are allowed. With a Co/Fe disorder, only one $A_{1g}$ mode is allowed for the $XX$ and $X'X'$ scattering configurations, but forbidden for $XY$. While the $E_{g}$ mode is allowed in $XX$, $X'X'$, but forbidden for $XY$ and $X'Y'$ scattering configurations. Our CFO film exhibits exhibits additional Raman modes of $A_{1g}$ symmetry centered at $473\;\mathrm{cm^{-1}}$ and $616\;\mathrm{cm^{-1}}$. The $E_{g}$ mode is allowed with $XX$, $X'X'$, and $X'Y'$, but forbidden for $XY$. The $F_{2g}$ modes are theoretically allowed with $XY$ and $X'X'$, but forbidden with $XX$ and $X'Y'$. Here again we observe an additional $F_{2g}$ mode at $657\;\mathrm{cm^{-1}}$. The issue of additional Raman peaks in the spectra of inverse spinels has been discussed in detail for closely related NiFe$_2$O$_4$ (NFO). The same arguments can also be applied to CFO as the structure and polarized Raman spectra of both materials are similar. Within the model proposed by Ivanov \textit{et al.},~\cite{ivanov_short-range_2010} the additional peaks reflect the fact that at a microscopic level Co$^{2+}$ and Fe$^{3+}$ are not randomly distributed, but ordered at the octahedral sites. These results show that our CFO films grown with the optimized deposition parameters exhibit the inverse spinel structure with indication of cation ordering at the octahedral sites.

\begin{figure}[h,b,t]
  \includegraphics[width=85mm]{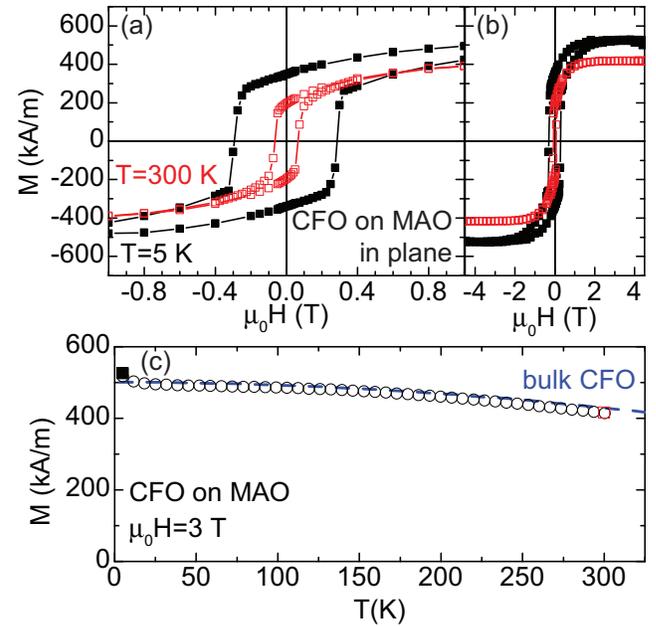}\\
  \caption[SQUID for CFO on MAO]{(Color online) SQUID magnetometry measurements of a CFO film grown on MAO substrate at $670\;^\circ\mathrm{C}$. (a) Hysteresis loops measured at $5\;\mathrm{K}$ (black, closed squares) and $300\;\mathrm{K}$ (red, open squares) with magnetic field applied along the [001] in-plane direction. At lower temperatures the coercive field increases. (b) Same hysteresis loops as in (a) at larger external magnetic fields. For $5\;\mathrm{K}$ the film saturates at fields $\geq 2\;\mathrm{T}$. (c) Temperature dependence of the saturation magnetization measured at $\mu_0 H=3\;\mathrm{T}$ applied in plane (black circles). The dashed blue line represents the temperature dependent saturation magnetization for bulk CFO taken from~\onlinecite{Pauthenet1952}. The black and red squares are the values extracted from the $M(H)$ loops at $5\;\mathrm{K}$ and $300\;\mathrm{K}$, respectively. The diamagnetic signal from the MAO substrate has been subtracted for this analysis.}
  \label{figure:SQUID}
\end{figure}

We analyzed the magnetic properties of a $1\;\mathrm{\mu m}$ thick CFO film grown on MAO at $670\;^\circ\mathrm{C}$ using SQUID-magnetometry. The obtained data has been corrected by subtracting the temperature independent, diamagnetic contribution from the MAO substrate. The hysteresis loops ($M(H)$) obtained at $5\;\mathrm{K}$ and $300\;\mathrm{K}$ are shown in Fig.~\ref{figure:SQUID}(a) and (b). The coercive field increases from $70\;\mathrm{mT}$ at $300\;\mathrm{K}$ to $300\;\mathrm{mT}$ at $5\;\mathrm{K}$. We attribute this change to an increase in magnetic anisotropy with decreasing temperatures. As evident from Fig.~\ref{figure:SQUID}(b) for $\mu_0 H\geq3\;\mathrm{T}$, the magnetization saturates at both temperatures. We extract a saturation magnetization of $420\;\mathrm{kA}$ at $T=300\;\mathrm{K}$ and $530\;\mathrm{kA}$ at $T=5\;\mathrm{K}$. The value for the saturation magnetization at $300\;\mathrm{K}$ is slightly lower than the already reported value for CFO thin films of $430\;\mathrm{kA}$ and the bulk value of $450\;\mathrm{kA}$.~\cite{dorsey_cofe2o4_1996,Coey2009,Goldman2006} As evident from the temperature dependence in Fig.~\ref{figure:SQUID}(c) and the $M(H)$ loops [Fig.~\ref{figure:SQUID}(b)] the saturation magnetization increases with decreasing temperature. We note that the stronger increase in saturation magnetization at $T\leq10\;\mathrm{K}$ is most likely caused by paramagnetic impurities in the MAO substrate. This effect also influences the quantitative value extracted for the saturation magnetization at $5\;\mathrm{K}$. Nevertheless, the increase in saturation magnetization is also present at $T\geq10\;\mathrm{K}$. Compared to the temperature dependence data of bulk CFO by Pauthenet~\cite{Pauthenet1952} [dashed blue line in Fig.~\ref{figure:SQUID}(c)], the film exhibits the same qualitative temperature dependence. The quantitative values for the film are for $T\geq10\;\mathrm{K}$ always slightly lower than the bulk data. The magnetic characterization reveals that our CFO films grown under optimized deposition conditions are state-of-the-art with a room temperature saturation magnetization that is $93\%$ of the bulk value.

\begin{figure}[h,b,t]
  \includegraphics[width=85mm]{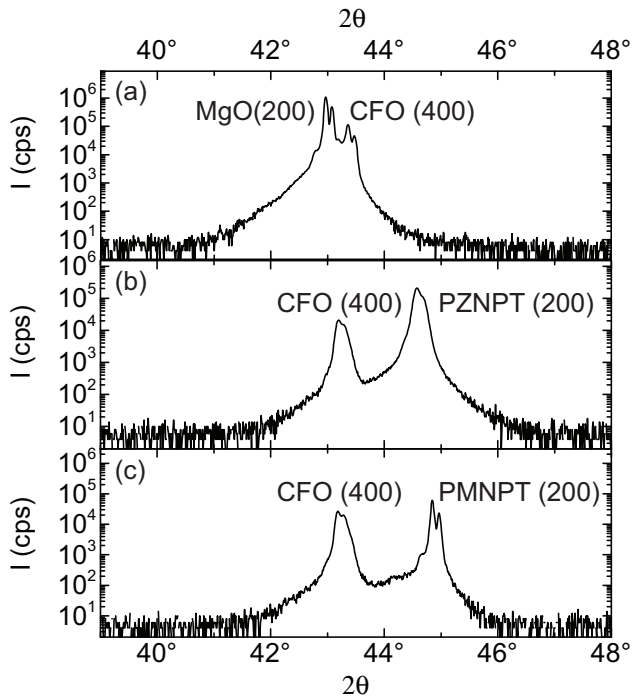}\\
  \caption[XRD spectra of CFO thin films on MgO,PMN-PT,PZN-PT]{XRD $2\theta-\omega$ scans for CFO films grown on (a) (100)-oriented MgO, (b) (100)-oriented PZN-PT, and (c) (100)-oriented PMN-PT using our optimized deposition conditions. All films exhibit a (100) orientation.}
  \label{figure:CFO_XRD_Substrates}
\end{figure}

Additionally, we have investigated the versatility of our optimized deposition conditions by growing CFO thin films on a variety of different substrates. For this study we deposited $1\;\mathrm{\mu m}$ thick CFO films on (100)-oriented MgO, along with piezoelectric lead magnesium niobate-lead titanate (PMN-PT) and lead zinc niobate-lead titanate (PZN-PT) substrates. We analyzed the structural properties using XRD. The results of the $2\theta-\omega$ scans are shown in Fig.~\ref{figure:CFO_XRD_Substrates}(a)-(c). For all of these different substrates we observe a epitaxial growth with the $(h00)$ planes parallel to the film surface. In addition, wide range $2\theta-\omega$ scans (not shown here) exhibit only reflections of the (100)-oriented CFO film and substrates. This excludes the existence of any secondary phases in the samples. From the XRD measurements we extracted the following CFO out-of-plane lattice constants: $8.340\;\mathrm{\AA}$ for MgO, $8.370\;\mathrm{\AA}$ for PZN-PT, and $8.374\;\mathrm{\AA}$ for PMN-PT; These differences can be attributed to the difference in lattice mismatch for the different substrates. For MgO we expect a tensile in-plane stress, while for PMN-PT and PZN-PT we expect an in-plane compressive stress. Only for CFO on MgO our experimental values and the expected stress state agree. As discussed above for the growth on MAO substrates, these results suggest that the density of misfit dislocations is the dominating factor, which determines the out-of-plane lattice constant. Moreover, we used $\varphi$-scans on the CFO film and substrate (220)-reflections to determine the in-plane epitaxial relationship (We here neglected the small rhombohedral distortion for PMN-PT and PZN-PT). For all three substrates we find: $\mathrm{MgO,PZN-PT,PMN-PT} (100)\:[001] \parallel \mathrm{CFO} (100)\:[001]$. 
The XRD analysis of the growth of CFO on different substrates using DLI-CVD shows that our optimized deposition parameters allow the growth of epitaxial (100)-oriented films independent of the substrate.

The surface roughness of the CFO films on different substrates has been evaluated by AFM scans on a $10\;\mathrm{\mu m}\times 10\;\mathrm{\mu m}$ area. From these measurements we obtain a RMS roughness of $3.9\;\mathrm{nm}$, $6.0\;\mathrm{nm}$, and $2.4\;\mathrm{nm}$ for our CFO films grown on MgO, PMN-PT, and PZN-PT, respectively. 

\section{Summary}
\label{Summary}
In summary, we have grown CFO films on (100)-oriented MAO substrates by DLI-CVD. A narrow deposition temperature window is found for growing high quality, epitaxial CFO films. XRD and TEM analysis of CFO films grown under these conditions yield the epitaxial relation: $\mathrm{MAO} (100)\:[001] \parallel \mathrm{CFO} (100)\:[001]$. AFM and SEM measurements show that the films exhibit a low RMS surface roughness of $1.1\;\mathrm{nm}$. Raman studies of the optimized films confirm the excellent structural quality and suggest an ordering of the Co$^{2+}$ and Fe$^{3+}$ ions on the B-lattice sites. The saturation magnetization of $420\;\mathrm{kA/m}$ at $300\;\mathrm{K}$ is very close to the CFO bulk value, and the temperature dependence of the saturation magnetization nicely agrees with measurements on bulk CFO. Moreover, we have demonstrated that the DLI-CVD process is highly versatile and enables the growth of epitaxial and smooth CFO films on a variety of other substrates, including MgO, PMN-PT and PZN-PT. A comparison of the expected in-plane strain and the observed out-of-plane lattice spacing support the fact that the density of misfit dislocations is the primary factor influencing the lattice constant of CFO. These results pave the way for the fabrication of magnetoelectric heterostructures and novel spintronic devices based on DLI-CVD grown CFO films.

\section{Acknowledgements}
This work was supported by NSF ECCS Grant No. 1102263. The work at the University of Houston was supported by the State of Texas through the Texas Center for Superconductivity (TcSUH).
\end{document}